# Comparison of Signaling and Media Approaches to Detect VoIP SPIT Attack

Ahmed Fawzy Gad
Information Technology Depr.
Faculty of Computers and Information
Menofia University, Egypt
ahmed.fawzy@ci.menofia.edu.eg

*Abstract*—IP networks became the most dominant type of information networks nowadays. It provides a number of services and makes it easy for users to be connected. IP networks provide an efficient way with a large number of services compared to other ways of voice communication. This leads to the migration to make voice calls via IP networks. Despite the wide range of IP networks services, availability, and its capabilities, there still a large number of security threats that affect IP networks and for sure affecting other services based on it and voice is one of them. This paper discusses reasons of migration from making voice calls via IP networks and leaving legacy networks, requirements to be available in IP networks to support voice transport, and concentrating on SPIT attack and its detection methods. Experiments took place to compare the different approaches used to detect spam over VoIP networks.

## I. INTRODUCTION

The Public Switched Telephone Network (PSTN) is a traditional telecommunications system, dedicated to voice exchange only. PSTN hardware and software are employed to only transmit voice, providing a very limited number of features. There are many challenges to add additional services to PSTN, making users, bounded to just make calls with some basic features. Moreover, due to the high cost of PSTN infrastructure, it is not easy to create a voice network that spans to large distances.

Due to the aforementioned limitations and not coping with the recent technological advances, the PSTN is nearing the end of its product lifecycle. On the other side, the opportunities presented by IP networks are immense. IP networks are now available in every home with many benefits that users can take advantage of everyday. The migration towards an IP-based telecommunications system seems to be a feasible solution helping the service providers to cover their network infrastructures in a large scale, providing huge bandwidth, and reducing long-distance charges.

To fully understand how IP networks will replace PSTNs, it is better to compare these two types of networks in terms of the network devices, protocols, and other aspects, to make a clear distinction between the two of them.

The PSTN is mainly used to connect end user telephone devices with each other. More than one telephone is connected via a device called Private Branch Exchange (PBX) that creates an end office connecting multiple end users. PBX provides a number of lines to establish voice calls inside a single organization. These end offices are connected together using special types of switches that create the backbone of PSTN. These switches provide lines for external call establishment outside of its local organization.

IP networks have a similar structure to that of the PSTN. IP networks have end user devices which are regular computers. These devices are connected together using a switch to create a local network. To enable communication among different local networks, routers are used which are the backbone of IP networks.

Due to the similarity in the structure of the two types of networks (PSTN and IP), it is possible to converge from PSTN to IP networks. However, IP networks are established for data exchange and regular data has different characteristics than voice. IP networks support offline data transmission and voice is online. The voice needs no delay, regular data on IP networks are not delay sensitive. The voice needs higher priority than other data types. So IP network on its native format cannot afford voice transmission and seems to need some modifications.

The modifications to be created on IP networks to support voice transmission are classified into two categories: hardware modifications and software modifications.

The hardware modifications are applied to the devices and the infrastructure of the IP networks in order to make it compatible with PSTN. Examples of devices that actually exist in the IP networks are: computers, hubs, switches, bridges, routers and access points. These devices cannot by default support handling voice and need some modifications. These modifications can be summarized to two points: the first one is changing the behavior of an existing device to make it more compatible to transmit voice and the second one is adding new devices other than the existing ones to support voice. Devices like routers give voice equal priority to other data types. An example of changes to be made on such devices is changing that to prioritize voice more. However, there is a number of services that cannot be achieved by changing the device behavior. In that case, adding new devices may help. Examples of such devices that can be added to the IP networks are: IP telephones with Analog-to-Digital Converter (ADC), Digital-to-Analog Converter (DAC), registration server enabling the user to specify its current location to forward calls on, proxy server to help saving time of the end user devices from locating destination and makes it on its own, redirect server and other types of devices.

The software modifications are applied to the IP networks' protocols. These modifications are basically in the form of adding or changing the behavior of a protocol. However, changing the behavior of existing protocols like UDP, TCP and IP cannot make IP networks fully support voice. In that case, some protocols are required to be added. Such protocols include Session Initiation Protocol (SIP) [2], (Real-time Transport Protocol (RTP) [3], (Real-time Transport Control Protocol) RTCP [3], (Session Description Protocol) SDP [4], (Resource Reservation Protocol) RSVP [5] and (Session Announcement Protocol) SAP [6]. SIP is a protocol that can handle possible security attacks on voice transported over the IP networks. In this paper, an overview of the SIP protocol is given.

This paper focuses on SPIT detection methods and gives a general overview about the different ways used to detect such attack compared to previous works [1, 2, 14, and 15] that just discuss SPIT attack with little details about the different detection algorithms for that spam.

The paper is organized as follows. Section II covers SIP and different messages used to establish a call, section III discusses the popular VoIP attacks, section IV focuses on SPIT and how it is different from regular e-mail spam, section V covers the different detection methods, and finally section VI compares among these presented detection methods VII makes conclusion and VIII presents the future work.

## II. SIP

SIP is a signaling protocol used for call establishment between end users in order to open a channel for voice communication. It is an application layer protocol, which is text-based like the HTTP protocol, with a number of messages that are exchanged between the callers.

Fig. 1 shows the basic SIP call establishment between two callers, A and B via a proxy server. Caller A wants to reach caller B, so it sends an INVITE message. This message just notifies the other user that there is a call waiting for acceptance. The proxy server receives the message, searches for caller B location and then forwards the message to that location once it is found. Caller B receives the message, notifies caller A that the INVITE message was successfully received by sending a RINGING information message. Caller B accepts the call and notifies caller A that the call actually started by sending an OK acknowledgment message. Then, caller A notifies caller B that it also accepted the call establishment and ready to start the conversation by sending an ACK message. The conversation starts and a media stream is opened between callers A and B. Finally, the call ends when sending a BYE message from either caller. Once the BYE message got received by the other caller, it sends an ACK message to confirm call termination.

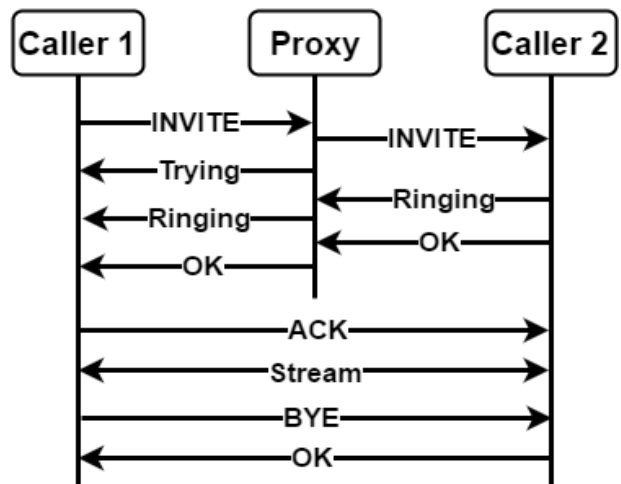

Fig. 1. SIP call establishment scenario.

As seen in Fig. 1, SIP messages are sometimes sent in the form of plain text. Due to that, VoIP protocols, such as SIP, can trigger attackers to make a compromise regarding the security of a particular SIP network.

## III. VoIP ATTACKS

This section illustrates some of identified threats/attacks in VOIP, their impact on the overall VOIP security.

Denial of Service (DoS) attack is an attempt to make a resource unavailable to its intended users. One common method is saturating the server with requests such that it cannot process legitimate requests. For example, assume that there is a server that can afford 100 users at a time; the attacker will send many messages to the server say 100 message, causing the server to be exhausted in processing and replying to these messages. In this case, there may exists a legal user that cannot be serviced from the server because its resources are fully used by the attack.

Man in the middle attack is a known attack in which there is a third party between the legal users connection. For a user to send a message to another one, it will pass through the attacker before reaching the destination. Attacker can drop the packet from reaching the other user or by changing message parameters and making the call last longer than its actual duration adding high costs than expected.

Registration hijacking is an attack where there is an attacker registered as a legal user so when a call get forwarded to that user it will be forwarded to the attacker too.

Spam over Internet Telephony (SPIT) is an attack where there is an attacker that sends spam calls to users connected to the Internet, as illustrated in Fig. 2. Next section gives more details about SPIT.

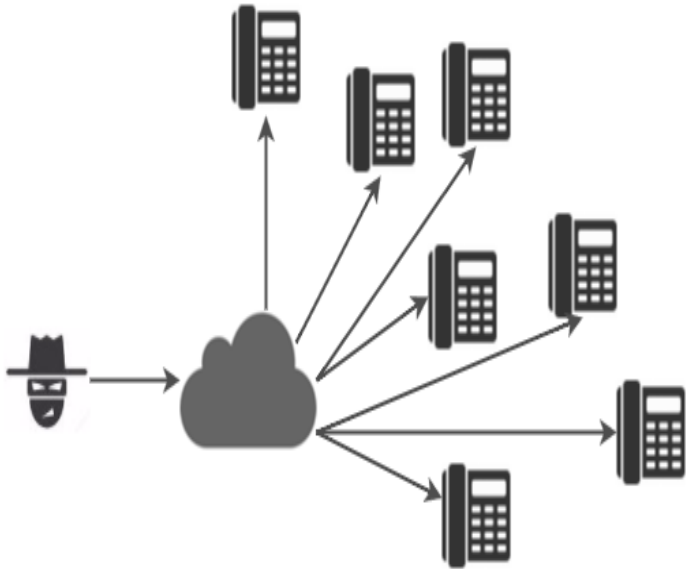

Fig. 2. Spam Over Internet Telephony (SPIT)

## IV. SPIT

SPIT or VoIP spam is one of the attack expected to have the major effect on the user experience. SPIT is an easy to propagate attack because it can be broadcasted to all IP phones connected to the Internet so attacker can record a voice message and create a multi-recipient call reaching more than one user at the same time. This increases the area of the attack and bandwidth wasted for propagating such spam calls. Detecting such attack is complex and challenging in some cases because it is not easy to trace the spam packets and have difficulties in finding useful information to be used to classify the call.

Both voice and e-mail have spams. However, there are some differences between voice and e-mail spams [16]. The e-mail spams and the SPIT spams are two different types of spam and both of them affect the user experience. The spam detector for e-mail is different than that of voice because of the different nature of e-mail and voice. A number of factors can be used in the comparison between both the email and the voice spam [9].

The first factor is the user interaction. Users get affected by e-mail less than voice spam because e-mail not interrupts user and stored away from regular user operations like being stored in a spam folder away from regular inbox. But voice spam directly reaches the user making bad experience.

The second factor is the detection complexity. E-mail spams can be easily detected because there is much information available for the spam detector. Spammer details are available in addition to the most important part which is the actual message content including its text, images, links, attachments, etc. This will allow the development of robust e-mail spam detectors. On the other side, for voice spam detector, at which only the caller details are known but the actual call content is unknown because it will only be available after accepting the call. This adds more challenges to voice spam detectors.

These and more are proofs that simply using spam detectors in e-mails to solve spam in VoIP is not the right solution [11].

## V. SPAM DETECTION IN VOIP

A general spam detector diagram is illustrated in Fig. 3. In this figure, the spammer tries to make a call to an Internet user. The spam detector extracts information from the call signaling messages to classify the call as spam or not. If it is spam, it will be rejected otherwise it will be accepted. Two types of information are used be used to classify the call to whether it is a spam or not: signaling and media [7].

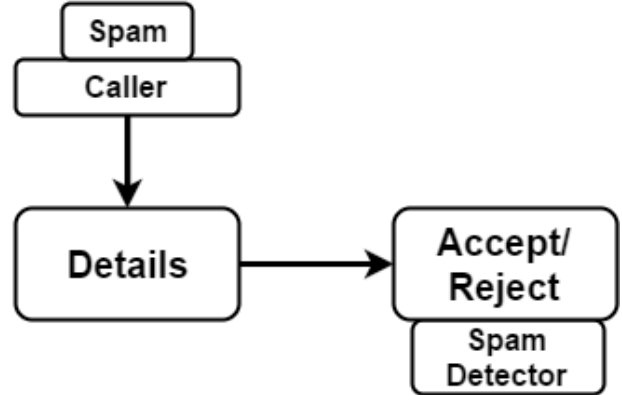

Fig. 3. Voice spam detector

SIP uses signaling information in the form of several messages to establish the call [8], [9]. There is also more than one step in the process of call establishment. So at which part of the call establishment spam should be detected to detect and stop it before starting affecting user experience. It is at the INVITE message. Fig. 4 shows an example of the INVITE message to better understand what its content [10].

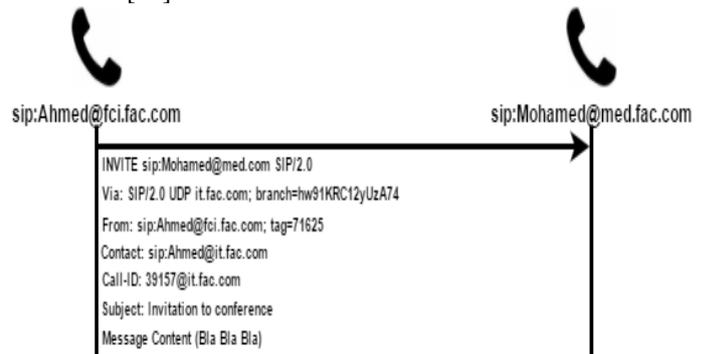

Fig. 4. SIP INVITE message

Important fields [17] of the INVITE message to be used in spam detection are: the *From* field that holds the spam address, the *Contact* field that holds future address, the *Call-ID* field, created by the spammer, the *Subject* field and the *Content-Type* field to specify the type of data to be transmitted. Fig. 5 shows the complete diagram of a spam detector using signaling information. If the spam detector classified the call as not spam, the call data will be forwarded to its destination.

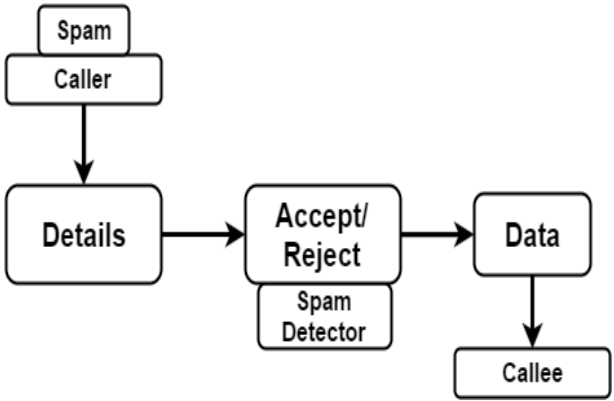

Fig. 5. Spam detection using signaling information

Using the signaling information is not an efficient way of spam detection because it is easy to misclassify the message as spam due to the lack of robust information in the time of spam detection process. So it is required to make more information available in time of spam detection.

Using media Information is a second approach of spam detection is which uses the call media content [18]. Like using e-mail actual content in detection of spam, actual call media can also be used in detecting voice spam [9] [11]. Fig. 6 shows how spam takes place and detected using the spam detector.

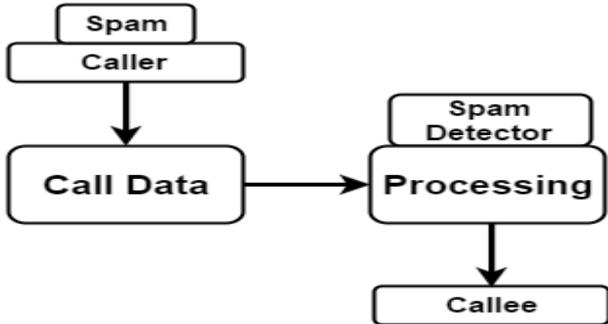

Fig. 6. Spam detection using call media information

As shown if Fig. 6, the spammer creates a call, the spam detector receives it and has to decide whether to accept it or reject it. However, in this case, there is extra information available for the classification process. The actual call data is now available. For each packet received by the spam detector, it is checked for being a spam. If it is spam, the call is dropped. This has the advantage of having a high percentage of knowing whether the call is spam or not because if it was a spam there will be some characteristics different from regular call like being a recorded message which will regularly not stops sending voice. However, there is a problem in this process. With each packet reaching the destination the spam detector must process it and make sure it is spam or not. Processing will last longer than usual because the data processed is voice and normally it takes much time to process such signals. So there will be a delay affecting end user experience. To eliminate this effect a modification to the normal process will be applied as shown in Fig. 7.

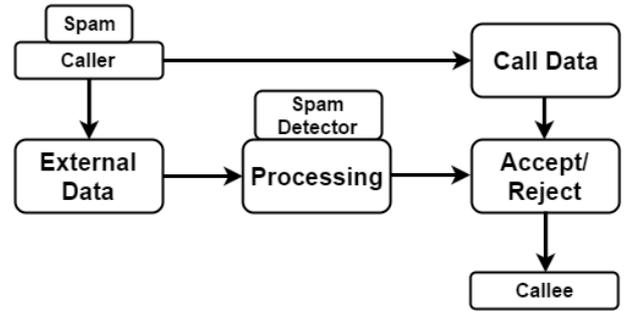

Fig. 7. Spam detection using call media information before actual call

Rather than processing each packet received, the spam detector will open a data stream between it and the caller for just a few seconds to make sure call is not a spam. Once caller sends its call to the spam detector, spam detector will not sends the call directly to the callee but using the created session will send some external data to the caller to know it is a spam or not based on certain behavior that a genuine caller makes [7], [9], [12]. Usually the caller will have some social behaviors that are different from the pre-recorded spam voice message [19] that helps in increasing the positive classification rate. This data can be a number of questions sent to the caller and spam detector receives its answers [5]. Based on the received data, spam detector can intelligently classify call as spam or not. If not spam then spam detector will allow call to be established between the end points directly. The drawback of this method is a bit delay before actual call establishment. But this delay can be acceptable in most cases because it occurs once in a call rather than with each packet [13].

## VI. COMPARISON

A comparison of the two main information source (signaling and media) used in SPIT attacks detection is shown in table 1. The metrics used in the comparison are early detection, detection complexity, speed, accuracy, and user experience. The early detection to SPIT is available only in signalling-based approach because processing signaling information is much simpler than media content. Media-based approach is more complex than signalling one because detection complexity increases as the amount of information required to be processed increases. In case of signalling-based approach, there are only a few headers to be checked compared to a large amount of media packets in media-based approach.

| Metric\Approach | Signaling | Media |
|---|---|---|
| Early detection | YES | NO |
| Detection Complexity | NO | YES |
| Fast | YES | NO |
| Accuracy | NO | YES |
| User Experience | YES | NO |

Table 1. Comparison between signaling and media approaches to SPIT detection

Signaling-based classifiers not consume too much time in their classification and it is easy to take the decision at early stages before affecting user experience but media-based approach requires the application of a signal processing algorithms to extract information useful in the classification. Usually signal processing is time-consuming. Media-based approach is more accurate compared to the signalling-based approach as more information is available and hence classification accuracy increases. Using media as the information source to the classifier gives high accuracy but unfortunately gives a bad user experience because most users do not need delay when establishing a call.

Tools used in the experiments are AsteriskNOW, Zoiper, WireShark, MySQL, and Python. Figure 8 presents the interaction among them.

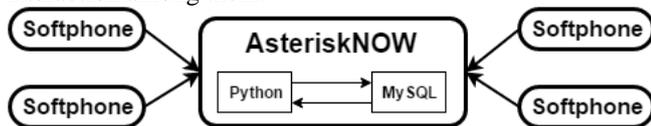

Fig. 8. VoIP testing environment

AsteriskNOW turns a computer into a communication server enabling VoIP services to be activated among clients. Zoiper creates softphones which are the clients sending and receiving voice calls simulating the actions of a real IP phone. WireShark is a network analyzer that captures packets sent or received through a network interface. MySQL is used to create a database holding details about spam calls. Python is the programming language used to create the firewall that will classify the coming calls as either spam or genuine. There are a number of Python modules used like PyShark and PyMySQL.

As a regular supervised classification system [20], there will be two major phases: training and testing. Supervised classification systems requires existing data in which the class of each entry is known.

The training data used in the experiments of system will created using an offline packet capture.

The target scenario is to have a session between two SIP devices, capture the packets, classify the call, and reject it if it was spam. This paper will have all of its experiments offline with no live packets captured. The reason of making the experiments offline is to make a restricted environment in which all packets transferred during the session are pre-stored and thus make it easy to create the training data because their class (spam or genuine) is already known. But in case of a live session there is no prior information about such received packets and we don`t know previously that the call is spam or not. Offline phase helps to assess the firewall accuracy.

The source of data in the offline phase is the WireShark packet sniffing tool. Some spam calls will be manually created between two clients and only the SIP and RTP messages will get captured using WireShark and get stored into a WireShark file for later use. There are some details get extracted from these messages based on whether the call classification uses signaling information or media.

**Classification Based on Signaling Information**

SIP is responsible for signaling and thus it is the protocol used to classify the call in this case. Just the INVITE message of the SIP will get analyzed for information to classify the call. The INVITE message contains many fields and just some selected fields will be used to classify the call [21, 22]. The fields selected are the ones that can be used to know the caller identity. Some fields are neglected because they don`t contain representative information about the caller.

The PyShark Python module helps reading the WireShark files and extracting fields used in the classification. It is used to read the WireShark packets, extract the SIP messages, and finally return all headers with their values.
Some of the fields to be used in the classification are SIP display info, caller IP address, SIP address, and via.
The caller IP address and the SIP address fields creates a URI of the form user@host. In spam calls it can be set to a value like Anonymous indicating that the call is spam like anonymous@anonymous.net. Whenever a value like that is found in the INVITE message the call is classified as spam and rejected.

Using the SIP display info header field the call can be classified as spam. Some spam calls will have their info set to some commercial values like company name. For example a company named TESTCOMPANY advertising for its products. The spam call generated from this company may hold some values like Summer Offer, Coming Soon, or whatever. It may also contains its name on this form t e s t c o m p a n y. It also may set that field to their site or mail address like t e s t c o m p a n y d o t c o m. Whenever an INVITE message contains the P-Asserted-Identity set to a value of these previous forms the call will be classified as spam and rejected.

When a new call is to be created, these fields will be extracted from the SIP INVITE message and based on them the call will be either classified as spam or genuine.

Using MySQL, a database is created holding some of the expected values found in the previous fields of the SIP INVITE message and marks the calls as spam. To connect Python to the database, the PyMySQL Python module is used to enable creating a database connection and exchanging data with the database.

When a new SIP INVITE message arrives, these fields will be extracted and a database search takes place. A binary number either one or zero will be returned reflecting whether the field was found in the database or not respectively. An AND operation will take place and if the result is 0 then the call will be classified as spam and got rejected.

**Classification Based on Media Information**

Using SIP information to classify the call is a very simple way to create a firewall because it doesn`t require any overhead to classify the call. Just compare fields with whatever stored in the database and make a decision. But unfortunately lack of information is a very critical point that degrade its accuracy in many situations. Also the classifier can be deceived in many cases by setting the SIP headers to some rational values that makes the call seem genuine.
So another approach to classify calls as either spam or genuine based on the actual media will be used. The data transferred during the call is the base for making a decision. The call will be accepted by default but got closed when the media is likely to be spam. The process of spam call detection using media information can be divided into two major phases. The first one is to access the media data and second one is making a decision based on such data. The

**following figure** summarizes the steps from receiving the packet until making a decision.

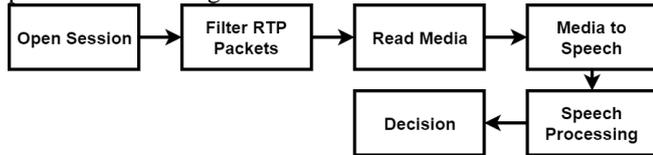

Fig. 9. Spam detection using media information

Previously, the SIP protocol was used to make a decision based on signaling information. This time the RTP protocol is the one to be used because it is responsible for carrying the voice signals.

The received packets will be filtered to get only the RTP packets. Part of each received RTP packet holds the media which is the target. But unfortunately the raw data can`t be applied to speech processing directly because it is a series of hexadecimals. So the raw data will be converted into speech signals which can be processed further. The goal is to know whether the speech signal belongs to an actual human talking in regular way or to a spammer. Based on some characteristics that capture the difference between them it is possible to classify the call correctly. In spam calls, some voice messages are pre-recorded and played directly after the call got accepted without making room for the other call participant to talk. Sometimes the opposite occurs as there are long silences during the call. Based on such characteristics, spam calls can be detected. Very simple speech signal features [23] will be extracted like zero crossing, signal mean, energy, and entropy to be able to differentiate spam from genuine calls.

Also if there is no silence then the call is likely to be spam. But it is not accurate to make a decision based on a single packet. So when there is no silences across multiple packets this indicates that the caller is not making room for the callee to participate in the call and this marks the call as spam.

Figure 10 presents genuine and spam speech samples used to train the classifier. Four samples per each class. The samples were selected to reflect the properties of the genuine and spam calls in which there are short silences in addition to not talking long time for genuine calls. For spam calls, samples were selected with long silences or no silences at all to reflect the behavior of spam calls. Table 2 shows the absolute mean extracted from each sample. Spam samples has very small or very large values with large deviation because they can have large silences and thus small values or no silences and thus high values. Genuine samples has in-between values.

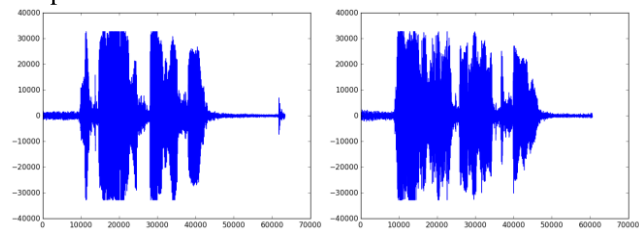

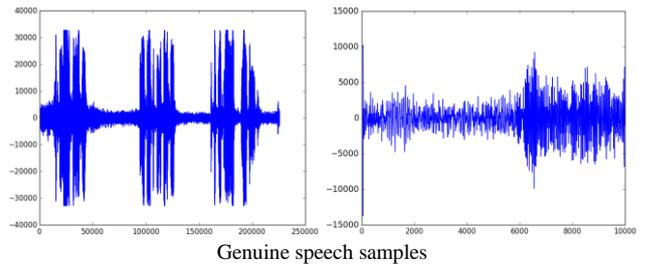

Genuine speech samples

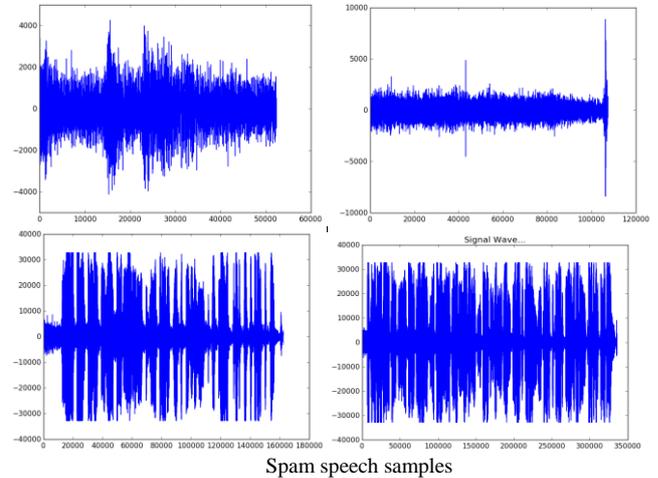

Spam speech samples

Fig. 10. Spam and genuine speech samples

Also table 2 shows the estimated time to process a speech sample data. The average time across all genuine and spam samples is 0.551 second. It is a very large time specially for real-time application like VoIP.

The time consumed when using the SIP signaling information is .2 second which is less than the time used by RTP speech data.

Table 2. Absolute Mean of Genuine and Spam Samples

| Genuine Samples | | | Spam Samples | | |
|---|---|---|---|---|---|
| # | Absolute Mean | Time | # | Absolute Mean | Time |
| 1 | 4.078 | 0.228 | 1 | 0.195 | 0.337 |
| 2 | 5.613 | 0.319 | 2 | 0.112 | 0.315 |
| 3 | 6.446 | 0.340 | 3 | 0.181 | 1.684 |
| 4 | 2.599 | 0.520 | 4 | 18.174 | 0.663 |

So each one has its pros and cons but it is possible to combine both of them and make benefit of each one to create a two-layer system to detect spam calls. In the first layer, the SIP signaling information will be used to classify the call. If it is spam then the call will be rejected with no further processing. If the call wasn`t classified as spam using SIP, then it will be accepted and applied to the second layer in which RTP speech data is processed.

When testing the system, a call is created and its RTP packets are fetched and then its data is converted into speech signal to extract the absolute mean. The absolute mean is then compared to values presented in the table 2 and got classified with the class of highest match.

## VII. CONCLUSION

The paper presented a review about voice transmission approaches starting by PSTN then IP networks with their components, devices, relationships, limitations of PSTN and how IP networks can solve these limitations by adding a number of extra features. There are many security threats in IP networks and one of them is SPIT. SPIT can be detected by two main approaches which are signaling and media content. Media-based approach is the most common one to detect and prevent SPIT.

## VIII. FUTURE WORK

After creating the training data that enables the classifier able to detect spam calls, next is to test the system on actual calls. The actual live session will be created using the Pcapy Python module. It is responsible for listening to a network interface and capture packets sent or received over that interface. Figure 11 summarizes the work done to classify the call.

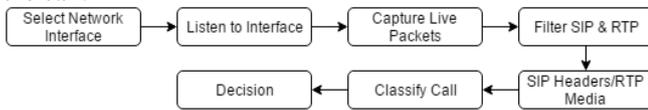

Fig. 11. Spam detection in live online session